\documentclass[aps,floats,prd,psfig,showpacs,twocolumn]{revtex4}
\input epsf
\usepackage{graphicx}% Include figure files

\newcommand{\beq}{\begin{equation}}
\newcommand{\beqa}{\begin{eqnarray}}
\newcommand{\eeq}{\end{equation}}
\newcommand{\eeqa}{\end{eqnarray}}

\newcommand{\lsim}{\lesssim}

\newcommand{\vect}[1]{\mbox{\boldmath${#1}$}}
\newcommand{\lmk}{\left(}
\newcommand{\rmk}{\right)}
\newcommand{\lnk}{\left\{ }

\newcommand{\rnk}{\right\} }
\newcommand{\lkk}{\left[}
\newcommand{\rkk}{\right]}
\newcommand{\lla}{\left\langle}
\newcommand{\p}{\partial}
\newcommand{\rra}{\right\rangle}

\newcommand{\vex}{{\vect x}}

\newcommand{\ven}{\vect n}

\newcommand{\vep}{{\vect p}}

\begin{document}
%\baselineskip 7mm
%\if0
%\draft
\title{Cosmological Constraints on the Very Low Frequency Gravitational-Wave Background} 
\author{Naoki Seto$^1$ and Asantha Cooray$^2$}
\affiliation{$^1$Theoretical Astrophysics, MC 130-33, California Institute of Technology, Pasadena,
CA 91125\\
$^2$Department of Physics and Astronomy, 4186 Frederick Reines Hall, University of California, Irvine, CA 92697
}
%\fi
\begin{abstract}
The curl modes of cosmic microwave background polarization
allow one to indirectly constrain the primordial background of gravitational waves with wavelengths roughly the
horizon size or larger with frequencies below 10$^{-16}$ Hz.  The planned high precision 
timing observations of a large sample of millisecond pulsars 
with the Pulsar Timing Array or with the Square Kilometer Array can either detect or constrain
the stochastic gravitational wave background at frequencies greater than roughly 0.1 years$^{-1}$. 
While there are no strong observational constraints on the gravitational wave background across six or more orders
of magnitude between 10$^{-16}$ Hz and 10$^{-10}$ Hz and it is difficult
 to get a constraint below $10^{-12}$Hz 
using objects in our Galaxy, we suggest that the anisotropy pattern of 
time variation of the redshift related to a sample of high redshift objects 
can be used to constrain the gravitational wave
background around 10$^{-12}$ Hz. Useful observations for the monitoring of an anisotropy signal in a global
redshift change include spectroscopic observations of the Ly-$\alpha$ 
forest in absorption towards a sample of quasars,
redshifted 21 cm line observations either in absorption or emission towards a sample of neutral HI
regions before or during reionization, and high frequency (0.1 Hz to 1 Hz) gravitational wave analysis 
of a sample of neutron star---neutron star binaries detected with gravitational wave instruments such as
the Decihertz Interferometer Gravitational Wave Observatory (DECIGO). The low frequency background can
also be constrained by arcsecond-scale anisotropy observations of the CMB. For reasonable observations in the
future involving extragalactic sources, we find best limits at the level of $\Omega_{\rm GW}  < 10^{-5}$ at a frequency around
10$^{-12}$ Hz while the eventual ultimate limit one cannot beat is $\Omega_{\rm GW}  < 10^{-11}$.
\end{abstract}
\maketitle

\section{introduction}
The observation of cosmic microwave background (CMB) anisotropies, especially using a map of the curl modes of polarization,
allows one to constraint the primordial background of gravitational waves with wavelengths roughly the
horizon size or larger with frequencies below 10$^{-16}$ Hz \cite{kamion}.  
Recent temperature anisotropy observations
at large angular scales with Wilkinson Microwave Anisotropy Probe (WMAP; \cite{bennet}) yields a limit (at the 2$\sigma$ level) of $\Omega_{\rm GW} < 5 \times 10^{-11}$ \cite{spergel,melch}
with a maximal sensitivity around 10$^{-17}$ Hz. 
Here, $\Omega_{\rm GW}$ is the fractional density contribution 
from a background of stochastic gravitational waves with the density $\rho_{\rm GW}$:
\begin{equation}
\Omega_{\rm GW}(f) = \frac{1}{\rho_c} \frac{d\rho_{\rm GW}}{d\log f} \, ,
\end{equation}
when the closure density is $\rho_c=3 H_0^2/8\pi G$. Here, $H_0$ is the Hubble constant. At these
super-horizon and horizon-size scales, a stochastic background of gravitational waves is expected from
inflationary physics \cite{kamion2,Maggiore:1999vm}. 
The direct detection of this signal is now considered to be one of the primary goals
of upcoming CMB polarization 
anisotropy observations both from ground and space with planned missions such as the
Inflation Probe of the NASA's Beyond Einstein Foundation Science Program.  Relative to the current
limit from intensity anisotropies, one can improve by
at least 5 orders of magnitude to the ultimate level allowed by the cosmic variance-limited removal of
a confusion related to lensing conversion of polarization modes \cite{seljak}. 
With a separation of lensing produced curl-modes, we expect the limit to improve down to
$\Omega_{\rm GW} < 10^{-16}$ with the best sensitivity again around a frequency around 10$^{-17}$ Hz \cite{knox}.
The proposed Explorer Probe of Inflationary Cosmology (EPIC), to be launched around
2012 as part of NASA's Beyond Einstein Foundation Science program, will be optimized to reach this level. 

When considering a subhorizon-scale wavelengths,  
the planned high precision timing observations of a large sample of
millisecond pulsars with the Pulsar Timing Array or with the Square Kilometer Array can either detect or 
constrain the stochastic gravitational wave background at frequencies greater than roughly $1/T_{\rm obs}$ 
where $T_{\rm obs}$ is the total observational duration of the pulsar
sample \cite{kramer}.   While the waves with wavelengths lower than $1/T_{\rm obs}$
 linearly change the observed interval of the pulse, the lower frequency limit of $1/T_{\rm obs}$ for the pulsar timing method 
comes from the fact that  one cannot distinguish their effect from  the long-term intrinsic
spin evolution  of pulsars.  Since observations are likely to be restricted to
less than 100 years, the pulsar timing arrays cannot constrain the gravitational wave 
background below $10^{-10}$ Hz. Observations of two millisecond pulsars over 17 years limit
the background to be $\Omega_{\rm GW} h^2 < 2 \times 10^{-9}$ (at the
95\% confidence level) \cite{lommen, thorsett},
while with the Pulsar Timing Array, using a larger sample of
pulsars, the limit is expected to reach the level of $2 \times 10^{-13}$
\cite{kramer}.
At these frequencies, the stochastic background of gravitational
waves is expected to be  primarily dominated by massive black hole binaries \cite{jaffe} 
and either a detection or a tight constrain
on the background is useful to understand the extent to which massive 
black holes merge at redshifts around unity.

Below frequencies probed by the millisecond pulsar timing
method, a background of stochastic waves is expected from models related to
global cosmic strings, phase transitions in the early universe, and
cosmic turbulence \cite{kosowsky}. Note also  that the primordial
background below $\sim 10^{-10}$Hz cannot be constrained by the
abundance of the light elements predicted by big bang nucleosynthesis \cite{Allen:1996vm}. 
At present, there are no strong observational constraints on the gravitational wave background 
across six or more orders of magnitude between 10$^{-16}$ Hz and
10$^{-10}$ Hz.   The
binary pulsars give a limit $\Omega_{\rm GW}h^2 < 
0.04$ for $10^{-11}{\rm Hz}<f<4.4\times10^{-9}{\rm Hz}$, and 
$\Omega_{\rm GW}h^2 <
0.5$ for $10^{-12}{\rm Hz}<f<10^{-11}{\rm Hz}$ from the time
variation of their orbital period 
compared with prediction of general relativity \cite{thorsett}. In
future they 
might give a limit at the $\Omega_{GW}h^2\sim 10^{-4}$ level that is determined
by uncertainties in the local acceleration of these binaries \cite{bertotti}. 
Using Galactic objects, it is difficult to  get information on the
background below $f\lsim 10^{-12}$ Hz as this corresponds roughly to the typical distance to galactic
objects. The longer waves act on  the target object and the observer in the same 
manner, and the measurement sensitivity  for these low frequency waves is strongly 
suppressed  due to a cancellation effect \cite{bertotti}. Therefore, it is essential that we use
 extra-Galactic objects to set a limit for the low frequency background below $10^{-12}$Hz.
The current  limit  comes from
observations of quasar proper motions across the sky (transverse motions) \cite{pyne}. For data collected over a 10 year span,
the resulting constraint on the stochastic background at a 
frequency below $2 \times 10^{-9}$ Hz is $\Omega_{\rm GW}h^2 < 0.11$
\cite{gwinn} (see also \cite{Jaffe:2004it} for prospects with the Square
Kilometer Array).

In this paper, to probe the background at frequencies below 10$^{-12}$ Hz, 
we propose measuring the anisotropy pattern of time variation of the redshift related to a sample of high redshift
objects.  Previously,  observations of  the time variation of the redshift have been
 proposed to study global cosmological parameters \cite{loeb}. 
While cosmology can induce either an acceleration or a deceleration to the redshift of an object,
the resulting change can be extracted through the monopole change from a sample
of objects spread over the sky since all sources are equally affected by cosmology.
The resulting change due to a gravitational wave background, at the lowest order in spherical moments, 
is present in the quadrupole moment of the time evolving redshift \cite{burke}.
 The use of the anisotropy pattern also
allows one to separate secondary effects that can cause a time-dependent redshift, such as
the peculiar acceleration of the observer which is present via a dipole at the lowest order.
We focus on the quadrupole measurement, but the measurement of additional 
multiple moments of the anisotropy can be used to increase the
confidence level of the derived limit. As we discussed,  the proposed method with using the time variation of
redshifts would be performed as a byproduct of the dark energy study,  in contrast to  the
method based on the transverse motions that is basically caused by perturbative effects. Therefore we will pay
special attention to relate the sensitivities of the measurement for the
dark energy and the gravitational wave background. As  we are studying
cosmology,  we can expect, in order of magnitude sense, that there is a
simple relation between their sensitivities that are properly normalized
by cosmological parameters such as $H_0$. It is, however, important to calculate
the relevant coefficients in discussing future prospects.
In fact, we find that  these coefficients are largely different from unity.

Here, extending the discussion in Ref.~\cite{loeb}, we consider a measurement of the quadrupole related to the anisotropy of the
time derivative of Ly-$\alpha$ forest redshifts seen in absorption towards a sample of quasars. Additional probes for the same purpose include 
low-frequency redshifted-21 cm observations either in absorption or emission towards a sample of neutral HI
regions before or during reionization \cite{Fie58}. 
Another possibility is the high frequency (0.1 Hz to 1 Hz) gravitational wave analysis of
of a sample of neutron star-neutron star binaries detected with gravitational wave instruments such as
the Decihertz Interferometer Gravitational Wave Observatory (DECIGO) \cite{Seto:2001qf}. These observations have been 
suggested for measurement of cosmological parameters based on the phase variation 
induced during the propagation of gravitational waves in a cosmologically expanding
background \cite{Seto:2001qf,takahashi}. While cosmologically induces a global phase shift, a low frequency
gravitational wave background induces an anisotropy to this phase shift and its presence can be
extracted from a sample of binaries spread over the sky with well-modeled wave forms. 
For reasonable future observations along these
lines, we find best constraints at the level of $\Omega_{\rm GW}  < 10^{-5}$ at a frequency around
10$^{-12}$ Hz, though, the eventual limit is probably at $\Omega_{\rm GW}  < 10^{-11}$. Reaching
this level is challenging as it requires high resolution observations of 21 cm background to
resolve all halos spatially with followup high resolution spectral measurements over a 10 year span over a bandwidth less
than a kHz.
In figure 1, we plot sensitivities of various current and future constraints 
for the low frequency gravitational wave background. From this figure, 
the importance of the extra-Galactic probes to study waves below $10^{-12}$ Hz, but above the regime studied by CMB experiments, is clear.

\begin{figure}[h]
 \begin{center}
 \epsfxsize=9.cm
 \begin{minipage}{\epsfxsize} \epsffile{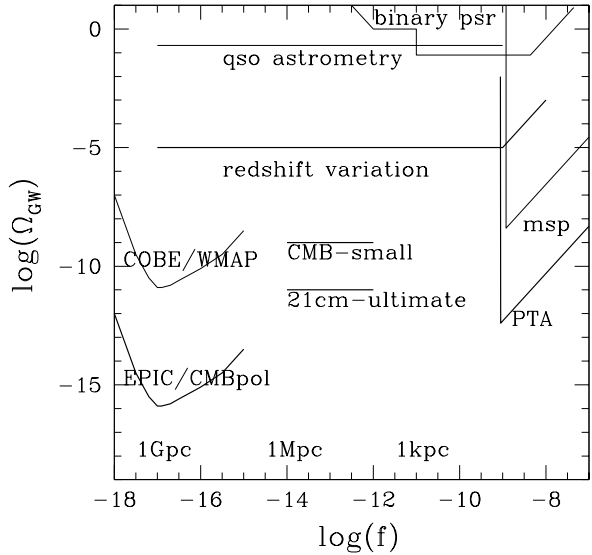} \end{minipage}
 \end{center}
\caption[]{ Sensitivities for various current (thin lines) and future
 (thick lines) measurements. 
%% AC: extended caption, better
Curves labeled 
``COBE/WMAP'' and ``EPIC/CMBpol'' show the current and future constraints based on CMB anisotropies at large
angular scales, while the limit labeled ``CMB-small'' is a potential constraint based on
arcsecond resolution, or multipole moments between $5 \times 10^4$ to $5 \times 10^6$,
observations of CMB anisotropies with a residual radio point source contamination consistent
with current anisotropy data. 
The curve labeled ``binary psr'' is the limit based on a sample of binary pulsars,
based on the time variation of their orbits, the curve labeled ``msp'' is the
current limit based on 17-year timing monitoring of 2 
millisecond pulsars while the curve labeled
``PTA'' is the limit based on proposed Pulsar Timing Array network. The limit based on ``QSO astrometry'' involves
proper motion observations of a sample of radio bright quasars while ``redshift variation'' involves
time monitoring of redshift variations in the Lyman-$\alpha$ forest
 towards bright quasars,  the 21 cm
line at very low radio frequencies and cosmological  neutron star
 binaries with ultimate DECIGO. The curve labeled ``21cm-ultimate'' is
 the eventual ultimate limit for method with 21 cm line.}  
\end{figure}

Incidently, one can also, in principle, use the monopole to constrain the energy density
of gravitational background. The resulting correction comes from the fact that a subhorizon-scale gravitational wave background
acts as an additional radiation field in the universe and the expansion rate of the universe is
modified to become $H^2(z)=H_0^2 \left[\Omega_m (1+z)^3 + (\Omega_R+\Omega_{\rm GW}) (1+z)^4 + \Omega_\Lambda\right]$,
for a spatially-flat universe with cosmological constant $\Omega_\Lambda$, the matter density
parameter $\Omega_M$ and the radiation content $\Omega_{R}$, all relative to the critical density.
Since matter and dark energy densities dominate the expansion rate over the redshift ranges
one can probe with current observational techniques, we do not pursue a limit based on this argument further.
An interesting constraint, however, could potentially be derived if one were to study earliest times in the universe
such as the era of big bang nucleosynthesis and using the number of massless neutrinos.
The constraint applies to waves with wavelengths
 below about a 10 pc, and the previous limit based on this argument constraints the high-frequency background, as observed today, to be
$\Omega_{\rm GW}  < 5 \times 10^{-4}$ \cite{Allen:1996vm} with a matter-radiation equality at a redshift of $\sim$ 3500 \cite{spergel}.
With improving measurements on the ratio of photon-to-baryons from both the cosmic microwave background anisotropy spectrum, with physics
dominating at the last scattering surface around $z \sim 1100$, and the same ratio measured independently based on
nucleosynthesis arguments, and can derive an improved constraint on the subhorizon gravitational background that  existed between
an age for the Universe of a minute to a few hundred thousand years. We will return to this topic in an upcoming paper.

Just as horizon-size waves produce CMB anisotropies, a gravitational wave background at frequencies
between 10$^{-14}$ to 10$^{-12}$ Hz can produce fluctuations in the CMB temperature at arcsecond angular scales.
Currently most CMB anisotropy observations probe down to  arcminute scales from acoustic peak structure at the degree scale, 
and prospects for even further small-scale CMB anisotropies observations are limited.
These observations are also likely to be challenging given 
additional secondary effects such as the Sunyaev-Zel'dovich effect \cite{SZ}. While most secondaries can be separated,
through properties such as the frequency spectrum relative to CMB \cite{cooray}, the foreground contamination from unresolved point-sources
is expected to be the biggest confusion. We make an estimate on the expected constraint by extrapolating the
foreground contamination seen in current anisotropy data at arcminute scales.

The discussion is organized as follows: In the next Section, we outline the calculation
related to an anisotropy pattern in the time evolving redshift of a sample of sources
in the presence of a gravitational wave background. In \S~III, we discuss various
cosmological probes related to this effect and estimate resulting limits that one
can potentially obtain in the future related to the presence of a low frequency, but sub-horizon,
 background of gravitational waves. We conclude with a summary in \S~IV.

\section{formulation}

We first study how a background of stochastic gravitational waves affect the redshift of a photon, or a massless particle, 
propagating to an observer today from a source at a high redshift. We consider
a spatially-flat Universe and write the metric that is perturbed by gravitational waves as 
\beq
ds^2=a(\eta)^2\lnk-d\eta^2+[\delta_{ij}+h_{ij}(\eta,\vex)]dx^i dx^j   \rnk,
\eeq
where $\eta$ is the conformal time and the scale factor $a(\eta)$ is normalized such that
$a(\eta_0)=1$ at the present epoch with $\eta=\eta_0$. We write the conformal
time interval and the normal time interval today as $\Delta \eta$  and $\Delta T$, 
respectively. Comparing to a background universe with $h_{ij}=0$, the propagation time of a photon changes by $\delta \eta$ due to the 
presence of gravitational waves. We can  formally write down $\delta \eta$ as a line integral along the photon path:
\beq
\delta \eta=\frac{n_i n_j}2\int_{\eta_0-L}^{\eta_0}
h_{ij}[\eta,(\eta-\eta_0)\ven] d\eta,
\eeq
where $\ven$ is the unit directional vector to the high redshift source and $L$ is its
comoving distance from the observer. This shift $\delta \eta$ will result in an apparent time
variation, $\Delta z$, to the redshift of the source, $z$, in a time interval $\Delta T$:
\beq
\frac{\Delta z}{\Delta T}=\frac{n_i n_j}2\int_{\eta_0-L}^{\eta_0}
\frac{\p h_{ij}}{\p \eta}[\eta,(\eta-\eta_0)\ven] d\eta.\label{change}
\label{eqn:dzdT}
\eeq
Since we are interested in wave lengths ($\lsim 10$Mpc) smaller than the horizon scale, 
 we can safely express the gravitational waves for our study as 
\beq
h_{ij}(f,\vep, \eta,\vex)=\exp[i2\pi f (\eta-\vep\cdot\vex)] D(\eta,f)
b_{ij}(f\vep) \label{wave},
\eeq
where $\vep$ is  the unit vector for the propagation direction of the
wave, $D(\eta,f)$ represents the cosmological evolution of the wave
amplitude and $b_{ij}(f\vep)$ is the random number for each wave
characterized by the vector $f\vep$.

We decompose eq.(\ref{wave}) as
\begin{eqnarray}
&&h_{ij}(f,\vep, \eta,\vex)=\exp[i2\pi f (\eta-\vep\cdot\vex)] \lnk
D(\eta,f)-D(\eta_0,f) \rnk \nonumber \\
&& \quad + \exp[i2\pi f (\eta-\vep\cdot\vex)] 
D(\eta_0,f)
b_{ij}(f\vep) \, ,
\label{dec}
\end{eqnarray}
and substitute this in Eq.~(\ref{change}). Our basic aim is to  extract 
effects that are common to many sources. Here, the contribution of the first
term in Eq.~(\ref{dec}) is negligible since  (i) we are considering
a wave with a wavelength shorter than the horizon scale, (ii) the common coherent
effect is made by the perturbation close to us, or the observer, and (iii) the  evolution
time  scale of the amplitude $({\dot D}/D)^{-1}$ is order of the age of the
Universe. As is well known \cite{det}, the second term in Eq.~(\ref{dec}) can be
expressed as a sum of two quantities, the information of the gravitational wave at the
source and at the observer. The former can be dropped for the same
reason as in the previous case. In summary, we can apply the standard
argument  for the low frequency gravitational background
by using  objects within our galaxy ({\it e.g.}  timing residual for a
galactic millisecond pulsar). 
 Random variations in $\Delta z/\Delta T$ for each
binary due to the gravitational wave background can be regarded as a
noise for measurement of the common signal. For these binaries, however, the expected magnitude of variation is 
smaller than other sources of noise ({\it e.g.} local  acceleration,
detector noise), and hereafter we neglect this noise component.

Following the calculation of Burke \cite{burke}, we obtain the  angular
pattern of the common redshift variation $\Delta z$ due to the background in the
time  interval 
$\Delta T$ as
\beq
\dot{z}_{GW}(\theta,\phi) \equiv \frac{1}{H_0}\lmk \frac{\Delta z}{\Delta T}\rmk_{GW}=
\sum_{l\ge 2}{\dot z}_{lm;GW}Y_{lm}(\theta,\phi)\label{ylm},
\eeq
where $\dot{z}_{GW}(\theta,\phi)$, given through Eq.~\ref{eqn:dzdT},
has the angular dependence of $(1-\cos \theta) \cos 2\phi$. This angular dependence can be
written as a combination of $Y_{l2}$ and $Y_{l-2}$, when $l \ge 2$
such that
\begin{equation}
(1-\cos \theta) \cos 2\phi = \sum_{l=2}^{\infty} (-1)^l \sqrt{\frac{(l-2)!}{(l+2)!}} \left[Y_{l2} + Y_{l-2}\right] \, .
\end{equation}

The angular power spectrum of coefficient ${\dot z}_{lmGW}$ is 
\beq
\lla {\dot z}_{lm;GW} {\dot z}_{l'm';GW} \rra=\frac{6 \Omega_{GW}
}{\pi (l+2) (l+1)l(l-1)} \delta_{ll'}\delta_{mm'} \, ,
\eeq
and the parameter  $\Omega_{GW}$ here is defined by the normalized energy
density $\Omega_{GW}(f)$ per unit log frequency interval as
\beq
\Omega_{GW}=\int^{(\Delta T)^{-1}}_{f_{cut}}\Omega_{GW}(f)\frac{df}{f}.
\eeq
We have explicitly included a lower frequency cut-off since, below a certain
frequency,  energy density of the gravitational wave background will 
be severely constrained by CMB polarization observations.

We also define the angular power spectrum $C_l$ that is useful for a statistical study as
\begin{eqnarray}
C_l &=& \frac{1}{2l+1}\sum_m  |{\dot z}_{lm}|^2 \nonumber \\
    &=& \frac{6 \Omega_{GW}}{\pi (l+2) (l+1)l(l-1)} \, .
\end{eqnarray}

 In eq.(\ref{ylm}) it is 
important to note that the effect of the gravitational  wave background starts form
quadrupole $l=2$ mode and independent on the source redshift $z$, while
the  redshift change $\Delta z$ due to the cosmic acceleration/deceleration is a monopole
 ($l=0$) effect and  depends on the redshift $z$ \cite{loeb} that  corresponds to the
coefficient ${\dot z}_{00}$  as
\begin{eqnarray}
&&{\dot z}_{00;acc}=\nonumber \\
&-&\sqrt{4\pi}\lnk \lkk
\Omega_M(1+z)+\Omega_R(1+z)^2+\Omega_\Lambda(1+z)^{-2}\rkk^{1/2}
-1\rnk \nonumber. 
%&& =4\sqrt{\pi}X(z)/H_0 \, .
%\equiv 4\sqrt{\pi}R
\end{eqnarray} 
We define a parameter $X$ that characterizes the
second order correction to the relation  between time intervals at the
observer, 
 $\Delta t$, and the source, $\Delta t_z$  due to effective acceleration
as
\beq
(1+z)\Delta t_z=\Delta t-X \Delta t^2+O(\Delta t^3) \, .
\eeq
The parameter $X$ can have a finite value due to the  cosmological
acceleration,   gravitational wave background or the local
acceleration. For each object we fit the parameter $X$ to statistically
extract various information.
For the cosmological acceleration we have 
\begin{equation}
X(z) = \frac{1}{2}\left[H_0 - \frac{H(z)}{1+z}\right] = \frac{H_0{\dot z}_{00;acc}}{4\sqrt\pi}\, , 
\end{equation}
when the Hubble parameter at redshift $z$ is $H(z)$. The parameter $X(z)$ 
can also be written as $X(z) = 0.5 [\dot{a}(0)-\dot{a}(z)]$, where the overdot represents
a derivative with respect to the proper time. At small redshifts, when $z << 1$, $X(z) = -0.5 q_0 z$
where $q_0$ is the global deceleration parameter, $\Omega_m -\Omega_\Lambda/2$. In addition 
to time variation in the redshift, for gravitational waves emitted by
the source, the 
resulting effect related to global dynamics is an additional change to
the phase of the Fourier transformed gravitational waveform of a chirping
binary  by 
an amount $ \propto f^{-13/3} X $ whose frequency dependence is largely
different from the intrinsic binary evolution predicted by the post Newtonian expansion. This phase correction has been used
as a probe of cosmological 
parameters in Ref.~\cite{Seto:2001qf,takahashi}. The anisotropy of the phase correction, or  anisotropy
measurement of $X$ from a large sample of binaries spread over the sky, can be used a probe of
large wavelength gravitational waves. 

We assume that the number of  the probes ({\it e.g} neutron star
binaries, QSO line-of-sights) is sufficient and the effects caused by a
finite number of such sources, such as Poisson fluctuations, is not important for measurement of $z_{lm}$ at  low $l$s.
We also introduce the ratio $R\equiv X(z)/H_0$ to relate the resolution of the acceleration
measurement, from which $X(z)$ is derived, and the resulting constraint on the low-frequency gravitational wave background.
In the case of the measurement related to the cosmic acceleration, in terms of the monopole
variation associated with $X(z)$, the redshift dependence of $R$ is important and we analyze probes binned over some finite width in
redshift.  From the same data,  but using higher order
anisotropies, we get the constraint for $\Omega_{GW}$. With 
a statistical relation between the measurement error of the spherical
harmonic coefficients, $z_{lm}$, in terms of an overall error in the power spectrum, we have
\beq
\Delta C_l=\sqrt{\frac{2}{2l+1}}\left[C_l + \sigma^2 \dot {z_{00}} \right] \, ,
\label{eqn:noise}
\eeq
where $\sigma^2 \dot {z_{00}}$ is the variance related to the monopole measurement.
The minimum energy density of the gravitational wave background measurable by any of the methods
discussed below can be obtained with
\begin{equation}
\sigma_{\Omega_{\rm GW}}^{-2} = \sum_{l} \frac{1}{(\Delta C_l)^2}\left(\frac{\partial C_l}{\partial \Omega_{\rm GW}}\right)^2 \, .
\end{equation}
To obtain the limit, under the null hypothesis of no gravitational wave background, we set
$C_l=0$ in Eq.~\ref{eqn:noise}. Since the measurement of the whole power spectrum related to
redshift derivatives, or $X(z)$, may not be achievable in the near future, we only concentrate on the variance
related to the quadrupole, $C_2$. Note that the uncertainty in this measurement is now determined
by the error to which the monopole can be established, with $\sigma^2 \dot {z_{00}}$, given
by the error in $R\equiv X(z)/H_0$.  In figure 2, we plot the angular
spectrum $C_l$ for $\Omega_{GW}$. We also put the measurement error for each $C_l$, if the  sensitivity for the monopole mode $R$ is
$10^{-3}$ for a given redshift shell.

\begin{figure}[h]
 \begin{center}
 \epsfxsize=9.cm
 \begin{minipage}{\epsfxsize} \epsffile{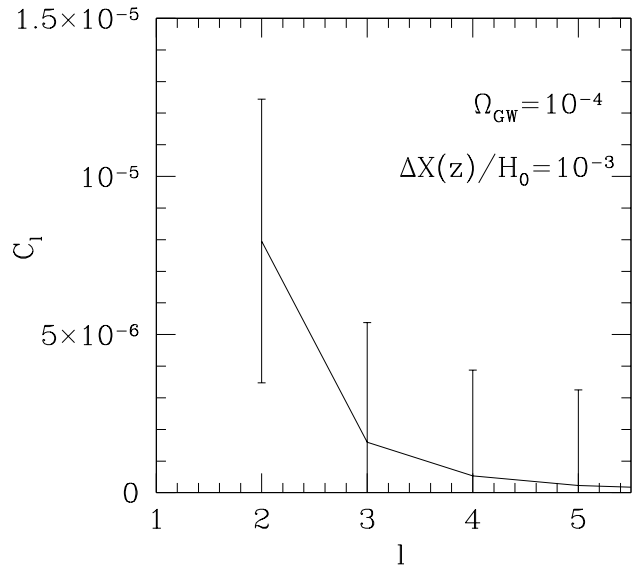} \end{minipage}
 \end{center}
\caption[]{ The angular power spectrum $C_l$ of a redshift shell due to the  low frequency  gravitational  wave
 background with $\Omega_{GW}=10^{-4}$. For the error bars we assume a
 case that we can measure the monopole mode $R$ with sensitivity
 $10^{-3}$ for the shell.}  
\end{figure}

Making use of the quadrupole anisotropy only, we obtain the 1-$\sigma$ constraint for the background as
\beq
\Omega_{GW} < 4.0\times 10^2 (\Delta R)_{shell}^2 \label{lim}
\eeq
from probes at each redshift shell, where $\Delta R$ is the error to which $R(z)$ can be established 
for such a shell. When we use the information of anisotropies up to $l=5$,  the
coefficient $4.0\times 10^2$ is 
reduced to $3.9\times 10^2$. As we commented earlier, the signal due to the gravitational wave
background does not depend on the redshift.  This means that the actual
constraint is better than the result using a single shell as one can combine shells
at different redshifts, independent of the exact redshift.  For cosmological
model with $h=0.7$, $\Omega_m=0.3$, and $\Omega_\Lambda=0.7$ the
magnitude of $R$ is $\sim 0.05$ around $z=1$. Therefore, if we want to
 study the evolution of the cosmic acceleration in a precious manner, we
need to have a resolution with an accuracy level $\Delta R\sim 0.005$, 
assuming a signal-to-noise ratio of 10, around a redshift
interval $dz \sim 0.1$. 
With this sensitivity for the redshift change, $\Delta z$, we can set the upper limit for
$\Omega_{GW}$ as
\beq
\Omega_{GW}\sim 4.0 \times 10^2 \times 10^{-1} \times (5 \times 10^{-3})^{2}\sim 10^{-3},
\eeq
where the factor $10^{-1}$ represents an average estimate on 
the improvement due to the increase in the  effective number of the shells.  If shells
between redshift of 1 and 3 can be divided at intervals of 0.1, this increase is a  factor of 
a few larger and one further improves the limit on $\Omega_{\rm GW}$.

\section{cosmological probes}

\subsection{Redshift variations in the line emission}

The use of time variation associated with the redshift of distant sources to measure global
cosmological parameters is described in Ref.~\cite{loeb}. The resulting change is an overall common shift
to the redshift distribution and the resulting change can be obtained by data separated in time.
For this purpose, the most useful observations are the ones that involve narrow lines either in emission or absorption
since the precise redshift can easily be ascertain from spectroscopic observations. In Ref.~\cite{loeb},
the use of the Lyman-$\alpha$ forest in absorption towards luminous quasars has been discussed.
The expected mean shift in the Lyman-$\alpha$ forest, at a redshift $\sim $ 3, is roughly 2 m s$^{-1}$ over a time
period of 100 years, and for presently favorable cosmological parameters. While this is a small number,
spectroscopic variations at the level of $\sim$ 1 m s$^{-1}$ are now routinely measured with 10 meter class
telescopes towards bright stars
in extra-Solar planetary searches using radial velocity measurements \cite{butler}. 
The advent of 30 meter class telescopes
over the next decade or more and the resulting increase in spectral resolution will improve searches
for small velocity shifts in high-redshift quasar spectra. 

While the precise measurement of the 
overall cosmological shift in a single quasar will require observations that
span many tens of years, a detection of the global change can be obtained through a statistical study of
a large number of quasars over a short time interval. While this will allow a measurement of the cosmological
parameters \cite{loeb}, the effect due to a gravitational wave background is high order and
requires a study related to the quadrupolar pattern of redshift shift.
While the use of a quadrupolar pattern allows one to separate the effect related to cosmological
acceleration, which is present in the monopole, and that due to local motion, present in the
dipole, with that expected from a background of gravitational waves,
the measurement, especially for an expected low energy density for the gravitational wave background is
harder. For $\Omega_{\rm GW}$ at the level of 10$^{-10}$ the anisotropy is at the level of
10$^{-5}$ to 10$^{-6}$, so the direct constrain one can put on $\Omega_{\rm GW}$ depends how well
the monopole can be established.  

Using estimates in Ref.~\cite{loeb} for observations of 1000
quasar sight lines with improved spectroscopic  data with a pixel scale of 0.5 km sec$^{-1}$) over a 10 year span,
and assuming no systematic or additional statistical errors such as in the calibration  of the spectrum,
 we find that one can  constrain the monopole to an accuracy of $\sim$ 1\%.  The resulting constrain on the low
frequency, with wavelengths out to $z \sim 3$, background of gravitational waves is at the level $\Omega_{\rm GW} < 10^{-3}$.
While the constrain is not significant, it is still useful to obtain this directly from the data as 
this will be one of the few methods to constrain the energy density of gravitational waves around $10^{-12}$ Hz.

In addition to the Lyman-$\alpha$ forest, one can constrain the energy density of gravitational waves
using the redshift distribution of 21 cm emitter. Here, the line emission is at low radio wavelengths
and statistics are expected to significantly  improve given the expected narrow width of these
lines, especially before the era of reionization, and the narrow band observations one can, in principle,
achieve at low radio frequencies. Expected statistics of HI emitters before reionization is not
well known and the complication here will be more related to foreground contamination \cite{cooray3} more than the
number of objects one can use to make these measurements. Improvements in foreground removal techniques 
suggest that the contamination can be mostly removed \cite{cooray4}.

We follow estimates in Ref.~\cite{pen} which  assume a model for
the cosmological distribution of neutral gas before reionization based on the halo approach \cite{CooShe02} and using so-called
``mini halos'' \cite{shapiro} that are beginning to
establish with adiabatically cooled gas with a temperature below that of the CMB thermal radiation at those
redshifts.  We estimate the line width of each signal to be around 3 km sec$^{-1}$ and to be dominated by
thermal motions within each halo. This compares to line widths of order 20 km sec$^{-1}$ in the Lyman-$\alpha$ spectrum.
Since the characteristic mass scale of each of these mini halos is around 10$^5$ M$_{\rm sun}$ to 10$^6$  M$_{\rm sun}$, one
expects a total of, in principle, 10$^{18}$ halos across the whole sky with a size of 1 kpc, or a projected size of
30 milliarcseconds. While the whole set of mini halos are not needed, using a sample of 10$^6$ brightest 21 cm lines 
over the whole sky, and assuming observations with an observational bandwidth of 0.2 kHz, which corresponds to a velocity of
$\sim$ 0.4 km sec$^{-1}$, we find that one can, in principle, constrain the background to be below
$\Omega_{\rm GW} < 10^{-5}$. The ultimate limit related to 21 cm anisotropies, using the whole sample of 10$^{18}$ halos and assuming sufficient
resolution to resolve them in redshift space and ignoring source confusion, 
is $\Omega_{\rm GW} < 10^{-11}$ with 10yr observation. This is comparable to the current limit from CMB 
anisotropies, but at frequencies between 10$^{-18}$ Hz and 10$^{-16}$ Hz.

\subsection{Gravitational waves from neutron star binaries around 1Hz}

Here we discuss the prospect of the measurement of
the background $\Omega_{GW}$ at the very low frequency regime ($\lsim 10^{-12}$Hz) with 
proposed  space gravitational wave missions, such as, the Big Bang Observer (BBO) or DECIGO,
whose optimum sensitivities are between LISA and ground based detectors
and  around $0.1$ Hz to 1 Hz. Note that this frequency range of 0.1 Hz to 1 Hz is largely
different from the band at $f \lsim 10^{-12}$Hz.

The most important aim of these detectors is the direct detection of the
gravitational wave background from early universe with $\Omega_{GW}\lsim
10^{-15}$ level at  $0.1$ to 1 Hz. For this measurement, the foreground gravitational  waves  from
various astrophysical sources must be subtracted from the data streams down to some
appropriate level. The cosmological  neutron star binaries are expected to produce a strong foreground,  and
we need an extensive fitting analysis with accurate templates to remove these
sources from the data. In this fitting procedure we can measure not only the intrinsic binary parameters,
but also  the total extrinsic effect made by the cosmic or
the local acceleration through the phase variation related to $X$. In the presence of a 
low frequency gravitational wave background, extracted $X$ values for a sample of binaries spread over the sky, over
a certain bin width in redshift, are suppose to vary with an anisotropic pattern.
As discussed, the quadrupolar pattern can in return be used as a probe of the low frequency gravitational wave background.
  Therefore cosmological neutron star binaries can be regarded as  probes of the dark energy, through the
monopole variation related to $X(z)$ \cite{Seto:2001qf,takahashi}, or the low frequency
gravitational wave background through the anisotropy pattern of $X$. While the discussion  is related to neutron star
binaries detectable in the 0.1 Hz to 1 Hz band,  we can also make a similar argument for
stellar mass black hole binaries in the same band, 
though their intrinsic waveforms would be generally  more complicated and the extraction of $X(z)$, through phase changes,
would be a complicated procedure. The local  acceleration might be
generally larger than the neutron star binaries.

At present, it is not clear how well we can actually resolve and subtract
the gravitational  wave contribution of each merging neutron star binary in BBO or DECIGO data streams.
For example, the confusion effect is more important for a binary with
a longer  time  before coalescence,  as its gravitational waves
 are in a lower frequency regime  where the number density of binaries per
unit frequency increases. In addition to binaries, the background made by other
sources such  as  supernovae could be a fundamental noise \cite{buonanno}. 

Now we study how well we might set the constraint for $\Omega_{GW}$ at
$f\lsim 10^{-12}$Hz based on the recent analysis for the measurement of
the cosmic acceleration $R\equiv X(z)/H_0$ \cite{takahashi}. Following their paper, we use the coalescence rate of neutron 
star binaries at $10^{-6}\rm /yr/Mpc^3$ assuming it is constant in time.
The total number of binaries in a redshift shell $z\sim z+dz$ is $\sim 10^{4}$/yr for a width $dz\sim 0.1$ at $z\sim 1$.
As we mentioned, in addition to the measurement error $\Delta X_{det}$
caused by  detector noises,  the random local acceleration becomes an effective
noise $\Delta X_{local}$ for estimating  the cosmic acceleration signal or the gravitational wave
background. In the case of a neutron star binary, its magnitude would correspond to
$\Delta X_{local}/H_0\sim 10^{-1}$ in terms of the noise $\Delta R$ for each 
binary \cite{binary}. Therefore, for a shell with a width $dz\sim 0.1$, the estimation of the common signal
 is limited by the fluctuations due to the local acceleration around
$(\Delta R)_{\rm shell}\sim 
10^{-1}\times \sqrt{10^{-4}}\sim 10^{-3}$ for an observation period of
order of one year. The final fluctuations after using all shells becomes
$\Delta R\sim(\Delta R)_{\rm shell}\times \sqrt{N_{\rm shell}}\sim 10^{-3.5}$. With an
effective number for shells of $N_{\rm shell}=10$, we can set a upper
limit of $\Omega_{GW}\sim5\times 10^{-5}$  using  the relation~(\ref{lim}). 

As for the measurement error $\Delta X_{det}$ determined by the detector
noises, we can reach the level $(\Delta R)_{\rm shell}\sim 3\times 10^{-4}$
(corresponding to $\Delta X_{det}/H_0\sim 3\times 10^{-2}$ for each binary) with a 3 year
integration by using the ultimate DECIGO (see figures 3 and 4 in Ref.~\cite{takahashi}). 
But with a one year  integration, the measurement error becomes $(\Delta R)_{\rm shell}\sim 3\times 
10^{-3}$  and dominates the fluctuation by the local acceleration. In
this estimation we did not include astrophysical confusion noises.

%% AC: New section
\subsection{Small Angular Scale CMB Anisotropies}

If a gravitational wave background exist at frequencies, say,
between 10$^{-14}$ Hz and 10$^{-12}$ Hz, the time evolution of the background would produce
a signature in CMB anisotropies through metric perturbations \cite{sachs}. This is similar to temperature fluctuations
produced at large angular scales with horizon-scale waves. In terms of
the angular power spectrum of CMB anisotropies, the signature would be
at multipole moments between $\sim 6 \times 10^4$ and $6 \times 10^6$. Such small-scale anisotropies,
at arcsecond angular scales, are not probed by any of the present observations. In fact,
no dedicated plans exist for the CMB anisotropy observations at such small scales though certain
radio interferometers planned for purposed other than CMB, such as the 
Atacama Large Millimeter Telescope (ALMA), can be used to make the required measurements. 

The fluctuations at
these scales are dominated by unresolved radio point sources, which produce a shot-noise 
spectrum. For comparison, assuming a radio point-source removal similar of the Cosmic Background Imager (CBI; \cite{mason}) experiment, where
residual fluctuations were at the level of 150 $\mu$K$^2$ in the smallest angular scale bin in
multipole moments between 2000 and 3500, we find that the resulting limit to CMB temperature
fluctuations in a bin with multipole moments between $\sim 6 \times 10^4$ and $6 \times 10^6$ is
$\Delta T < 3000 \mu$K. This results in a limit for the gravitational wave background between
10$^{-14}$ Hz and 10$^{-12}$ Hz of $\Omega_{\rm GW} < 10^{-9}$. To reach the ultimate level implied
by 21 cm observations require a factor of 100 improvement in foreground removal. Since foregrounds are
already restricted to a few percent level, this is a challenging task. 

\section{Summary}

The curl modes of cosmic microwave background polarization
allow one to indirectly constrain the primordial background of gravitational waves with wavelengths roughly the
horizon size or larger with frequencies below 10$^{-16}$ Hz.  The planned high precision 
timing observations of a large sample of millisecond pulsars 
with the Pulsar Timing Array or with the Square Kilometer Array can either detect or constrain
the stochastic gravitational wave background at frequencies greater than roughly 0.1 years$^{-1}$. 
While there are no strong observational constraints on the gravitational wave background across six or more orders
of magnitude between 10$^{-16}$ Hz and 10$^{-10}$ Hz, we suggest that by monitoring the anisotropy pattern of 
time variation of the redshift related to a sample of high redshift objects 
one can constrain the gravitational wave
background below 10$^{-12}$ Hz where methods using   Galactic objects
do not work well.  Direct measurement of the time variation of redshift
is one of the potential  method to get information of the dark
energy. Analyses in this paper would be helpful to discuss the prospects
of the low frequency gravitational wave background measurement in a
given observational condition of the dark energy study. 
 Useful observations for the monitoring of an anisotropy signal in a global
redshift change include spectroscopic observations of the Ly-$\alpha$ 
forest in absorption towards a sample of quasars,
redshifted 21 cm line observations, either in absorption or emission, towards a sample of neutral HI
regions before or during reionization, and high frequency (0.1 Hz to 1 Hz) gravitational wave analysis 
of a sample of neutron star---neutron star binaries detected with gravitational wave instruments such as
the Decihertz Interferometer Gravitational Wave Observatory (DECIGO). For reasonable observations in the
future, we find best limits at the level of $\Omega_{\rm GW}  < 10^{-5}$ at a frequency around
10$^{-12}$ Hz.

{\it Acknowledgments:}  This work has been
supported at Caltech by the Sherman Fairchild foundation, DOE DE-FG 03-92-ER40701 (AC) 
and NASA grant NNG04GK98G and the Japan Society for the Promotion of
Science (NS).


\begin{thebibliography}{99}

\bibitem{kamion}
M.~Kamionkowski, A.~Kosowsky and A.~Stebbins,
%``A probe of primordial gravity waves and vorticity,''
Phys.\ Rev.\ Lett.\  {\bf 78}, 2058 (1997)
[arXiv:astro-ph/9609132];
%%CITATION = ASTRO-PH 9609132;%%
U.~Seljak and M.~Zaldarriaga,
%``Signature of gravity waves in polarization of the microwave background,''
Phys.\ Rev.\ Lett.\  {\bf 78}, 2054 (1997)
[arXiv:astro-ph/9609169].
%%CITATION = ASTRO-PH 9609169;%%


%\cite{Bennett:2003bz}
\bibitem{bennet}
C.~L.~Bennett {\it et al.},
%``First Year Wilkinson Microwave Anisotropy Probe (WMAP) Observations:
%Preliminary Maps and Basic Results,''
Astrophys.\ J.\ Suppl.\  {\bf 148}, 1 (2003)
[arXiv:astro-ph/0302207].
%%CITATION = ASTRO-PH 0302207;%%


\bibitem{spergel}
D.~N.~Spergel {\it et al.}  [WMAP Collaboration],
%``First Year Wilkinson Microwave Anisotropy Probe (WMAP) Observations:
%Determination of Cosmological Parameters,''
Astrophys.\ J.\ Suppl.\  {\bf 148}, 175 (2003)
[arXiv:astro-ph/0302209].
%%CITATION = ASTRO-PH 0302209;%%

\bibitem{melch}
A.~Melchiorri and C.~J.~Odman,
%``The inflationary gravity waves in light of recent cosmic microwave
%background anisotropies data,''
Phys.\ Rev.\ D {\bf 67}, 021501 (2003)
[arXiv:astro-ph/0210606].
%%CITATION = ASTRO-PH 0210606;%%

\bibitem{kamion2}
M.~Kamionkowski and A.~Kosowsky,
%``The cosmic microwave background and particle physics,''
Ann.\ Rev.\ Nucl.\ Part.\ Sci.\  {\bf 49}, 77 (1999)
[arXiv:astro-ph/9904108].
%%CITATION = ASTRO-PH 9904108;%%

%\cite{Maggiore:1999vm}
\bibitem{Maggiore:1999vm}
M.~Maggiore,
%``Gravitational wave experiments and early universe cosmology,''
Phys.\ Rept.\  {\bf 331}, 283 (2000)
[arXiv:gr-qc/9909001].
%%CITATION = GR-QC 9909001;%%


%\cite{Zaldarriaga:1998ar}
\bibitem{seljak}
M.~Zaldarriaga and U.~Seljak,
%``Gravitational Lensing Effect on Cosmic Microwave Background Polarization,''
Phys.\ Rev.\ D {\bf 58}, 023003 (1998)
[arXiv:astro-ph/9803150];
%%CITATION = ASTRO-PH 9803150;%%
A.~Cooray and M.~Kesden,
%``Weak Lensing of the CMB: Extraction of Lensing Information from the
%Trispectrum,''
New Astron.\  {\bf 8}, 231 (2003)
[arXiv:astro-ph/0204068];
%%CITATION = ASTRO-PH 0204068;%%
M.~Kesden, A.~Cooray and M.~Kamionkowski,
%``Lensing Reconstruction with CMB Temperature and Polarization,''
Phys.\ Rev.\ D {\bf 67}, 123507 (2003)
[arXiv:astro-ph/0302536].
%%CITATION = ASTRO-PH 0302536;%%



%\cite{Knox:2002pe}
\bibitem{knox}
L.~Knox and Y.~S.~Song,
%``A limit on the detectability of the energy scale of inflation,''
Phys.\ Rev.\ Lett.\  {\bf 89}, 011303 (2002)
[arXiv:astro-ph/0202286];
%%CITATION = ASTRO-PH 0202286;%%
M.~Kesden, A.~Cooray and M.~Kamionkowski,
%``Separation of gravitational-wave and cosmic-shear contributions to  cosmic
%microwave background polarization,''
Phys.\ Rev.\ Lett.\  {\bf 89}, 011304 (2002)
[arXiv:astro-ph/0202434].
%%CITATION = ASTRO-PH 0202434;%%

\bibitem{kramer}
M.~Kramer,
%``Fundamental Physics with the SKA:Strong-Field Tests of Gravity Using Pulsars
%and Black Holes,''
arXiv:astro-ph/0409020;
%%CITATION = ASTRO-PH 0409020;%%
G.~Hobbs,
%``Pulsars and gravitational wave detection,''
arXiv:astro-ph/0412153.
%%CITATION = ASTRO-PH 0412153;%%




\bibitem{lommen}
A.~N.~Lommen,
%``New limits on gravitational radiation using pulsars,''
arXiv:astro-ph/0208572;
%%CITATION = ASTRO-PH 0208572;%%
V. M.~Kaspi, J. H.~Taylor  and  M. F. Ryba, Astrophys.\ J.\  {\bf 428},
 713 (1994).

\bibitem{thorsett}
S. E. Thorsett and R. J. Dewey,
Phys.\ Rev.\ D {\bf 53}, 3468 (1996).


\bibitem{jaffe}
A.~H.~Jaffe and D.~C.~Backer,
%``Gravitational waves probe the coalescence rate of massive black hole
%binaries,''
Astrophys.\ J.\  {\bf 583}, 616 (2003)
[arXiv:astro-ph/0210148].
%%CITATION = ASTRO-PH 0210148;%%


\bibitem{kosowsky}
A.~Kosowsky and M.~S.~Turner,
%``Gravitational radiation from colliding vacuum bubbles: envelope
%approximation to many bubble collisions,''
Phys.\ Rev.\ D {\bf 47}, 4372 (1993)
[arXiv:astro-ph/9211004];
%%CITATION = ASTRO-PH 9211004;%%
M.~Kamionkowski, A.~Kosowsky and M.~S.~Turner,
%``Gravitational radiation from first order phase transitions,''
Phys.\ Rev.\ D {\bf 49}, 2837 (1994)
[arXiv:astro-ph/9310044];
%%CITATION = ASTRO-PH 9310044;%%
A.~D.~Dolgov, D.~Grasso and A.~Nicolis,
%``Relic backgrounds of gravitational waves from cosmic turbulence,''
Phys.\ Rev.\ D {\bf 66}, 103505 (2002)
[arXiv:astro-ph/0206461].
%%CITATION = ASTRO-PH 0206461;%%

%\cite{Allen:1996vm}
\bibitem{Allen:1996vm}
B.~Allen,
%``The stochastic gravity-wave background: Sources and detection,''
arXiv:gr-qc/9604033.
%%CITATION = GR-QC 9604033;%%



\bibitem{bertotti}
B. Bertotti, B. J. Carr and M. J. Rees, 
Mon. Not. R. Aston. Soc. {\bf 203}, 945 (1983).

\bibitem{pyne}
T.~Pyne, C.~R.~Gwinn, M.~Birkinshaw, T.~M.~Eubanks and D.~N.~Matsakis,
%``Gravitational radiation and very long baseline interferometry,''
Astrophys.\ J.\  {\bf 465}, 566 (1996)
[arXiv:astro-ph/9507030].
%%CITATION = ASTRO-PH 9507030;%%

\bibitem{gwinn}
C.~R.~Gwinn, T.~M.~Eubanks, T.~Pyne, M.~Birkinshaw and D.~N.~Matsakis,
%``Quasar proper motions and low-frequency gravitational waves,''
Astrophys.\ J.\  {\bf 485}, 87 (1997)
[arXiv:astro-ph/9610086].
%%CITATION = ASTRO-PH 9610086;%%


%\cite{Jaffe:2004it}
\bibitem{Jaffe:2004it}
A.~H.~Jaffe,
%``Observing Gravitational Radiation with QSO Proper Motions and the SKA,''
New Astron.\ Rev.\  {\bf 48}, 1483 (2004)
[arXiv:astro-ph/0409637].
%%CITATION = ASTRO-PH 0409637;%%




%\cite{loeb}
\bibitem{loeb}
A.~Loeb, 
Astrophys.\ J. Lett.  {\bf 499}, 111 (1998).



%\cite{burke}
\bibitem{burke}
W. L. ~Burke,
Astrophys.\ J.  {\bf 196}, 329 (1975)

\bibitem{Fie58} G. B. Field, Proc. IRE, {\bf 46}, 240 (1958);
 G. B. Field, Astrophys.\ J.\ {\bf  129}, 525 (1959).


%\cite{Seto:2001qf}
\bibitem{Seto:2001qf}
N.~Seto, S.~Kawamura and T.~Nakamura,
%``Possibility of direct measurement of the acceleration of the universe  using
%0.1-Hz band laser interferometer gravitational wave antenna in  space,''
Phys.\ Rev.\ Lett.\  {\bf 87}, 221103 (2001)
[arXiv:astro-ph/0108011].
%%CITATION = ASTRO-PH 0108011;%%



%\cite{takahashi}
\bibitem{takahashi}
R.~Takahashi and T.~Nakamura,
%``Determination of the equation of the state of the Universe using ~ 0.1 Hz
%Gravitational Wave Detectors,''
arXiv:astro-ph/0408547.
%%CITATION = ASTRO-PH 0408547;%%


\bibitem{SZ}
R.~A.~Sunyaev and Ya.~B.~Zel'dovich, Mon.\ Not.\ Roy.\ Astron.\ Soc.\ {\bf 190}, 413 (1980).

\bibitem{cooray}
A.~Cooray, W.~Hu and M.~Tegmark,
%``Large-Scale Sunyaev-Zel'dovich Effect: Measuring Statistical Properties with
%Multifrequency Maps,''
Astrophys.\ J.\  {\bf 540}, 1 (2000)
[arXiv:astro-ph/0002238].
%%CITATION = ASTRO-PH 0002238;%%


\bibitem{det}
S. Detweiler,  Astrophys.\ J.\ {\bf  1234}, 1100 (1979);
B. Mashhoon, Mon. Not. R. Aston. Soc. {\bf 199}, 659 (1982).


%\bibitem{cooray2}
%A.~Cooray, A.~J.~Farmer and N.~Seto,
%``Optical Identification of Close White Dwarf Binaries in the LISA Era,''
%Astrophys.\ J.\  {\bf 601}, L47 (2004)
%[arXiv:astro-ph/0310889].
%%CITATION = ASTRO-PH 0310889;%%

\bibitem{butler}
R.~P.~Butler {\it et al.},
%``Ultra-high-precision velocity measurements of oscillations in alpha Cen A,''
Astrophys.\ J.\  {\bf 600}, L75 (2003)
[arXiv:astro-ph/0311408].
%%CITATION = ASTRO-PH 0311408;%%


\bibitem{cooray3}
S.~P.~Oh and K.~J.~Mack,
%``Foregrounds for 21cm Observations of Neutral Gas at High Redshift,''
Mon.\ Not.\ Roy.\ Astron.\ Soc.\  {\bf 346}, 871 (2003)
[arXiv:astro-ph/0302099];
%%CITATION = ASTRO-PH 0302099;%%
A.~R.~Cooray and S.~R.~Furlanetto,
%``Free-Free Emission at Low Radio Frequencies,''
Astrophys.\ J.\  {\bf 606}, L5 (2004)
[arXiv:astro-ph/0402239];
%%CITATION = ASTRO-PH 0402239;%%
T.~Di Matteo, B.~Ciardi and F.~Miniati,
%``The 21 centimeter emission from the reionization epoch: extended and point
%source foregrounds,''
arXiv:astro-ph/0402322.
%%CITATION = ASTRO-PH 0402322;%%

\bibitem{cooray4}
M.~G.~Santos, A.~Cooray and L.~Knox,
%``Multifrequency Analysis of 21 cm fluctuations From the Era of
%Reionization,''
arXiv:astro-ph/0408515;
%%CITATION = ASTRO-PH 0408515;%%
X.~M.~Wang, M.~Tegmark, M.~Santos and L.~Knox,
%``Twenty-one centimeter tomography with foregrounds,''
arXiv:astro-ph/0501081.
%%CITATION = ASTRO-PH 0501081;%%


\bibitem{pen}
U.-L. Pen, New Astron.\, [arXiv:astro-ph/0305387].

%\cite{Cooray:2002di}
\bibitem{CooShe02}
A.~Cooray and R.~Sheth,
%``Halo models of large scale structure,''
Phys.\ Rept.\  {\bf 372}, 1 (2002)
[arXiv:astro-ph/0206508].
%%CITATION = ASTRO-PH 0206508;%%

\bibitem{shapiro}
H.~Martel, P.~R.~Shapiro, I.~T.~Iliev, E.~Scannapieco and A.~Ferrara,
%``On the Detectability of the Cosmic Dark Ages: 21-cm Lines from Minihalos,''
AIP Conf.\ Proc.\  {\bf 666}, 85 (2003)
[arXiv:astro-ph/0302335].
%%CITATION = ASTRO-PH 0302335;%%

\bibitem{buonanno}
A.~Buonanno, G.~Sigl, G.~G.~Raffelt, H.~T.~Janka and E.~Muller,
%``Stochastic gravitational wave background from cosmological supernovae,''
arXiv:astro-ph/0412277.
%%CITATION = ASTRO-PH 0412277;%%

\bibitem{binary}
T. Damour and J. H.~Taylor, Astrophys.\ J.\  {\bf 366}, 501 (1991).



\bibitem{sachs}
R.~K.~Sachs and A.~M.~Wolfe, Astrophys.\ J.\  {\bf 147}, 73 (1967).

\bibitem{mason}
B.~S.~Mason {\it et al.},
%``The Anisotropy of the Microwave Background to l = 3500: Deep Field
%Observations with the Cosmic Background Imager,''
Astrophys.\ J.\  {\bf 591}, 540 (2003)
[arXiv:astro-ph/0205384].
%%CITATION = ASTRO-PH 0205384;%%


\end{thebibliography}
\end{document}